\documentclass[twoside]{article}
\usepackage{amsmath, graphicx, cite}
\usepackage[a4paper, total={7in, 9in}]{geometry}
\usepackage{multicol}
\usepackage{url}
\usepackage{hyperref}
\usepackage{cite}

\begin{document}

\title{Attention-guided Spectrogram Sequence Modeling with CNNs for Music Genre Classification}
\author{Aditya Sridhar \\ Department of Computer Science, IIT Hyderabad \\ \texttt{es21btech11004@iith.ac.in}}
\date{30 October 2024}

\maketitle

\begin{abstract} Music genre classification is a critical component of music recommendation systems, generation algorithms, and cultural analytics. In this work, we present an innovative model for classifying music genres using attention-based temporal signature modeling. By processing spectrogram sequences through Convolutional Neural Networks (CNNs) and multi-head attention layers, our approach captures the most temporally significant moments within each piece, crafting a unique "signature" for genre identification. This temporal focus not only enhances classification accuracy but also reveals insights into genre-specific characteristics that can be intuitively mapped to listener perceptions. Our findings offer potential applications in personalized music recommendation systems by highlighting cross-genre similarities and distinctiveness, aligning closely with human musical intuition \cite{moore2012introduction}. This work bridges the gap between technical classification tasks and the nuanced, human experience of genre. \end{abstract}

\begin{multicols}{2}

\section{Introduction} The task of music genre classification is complex, requiring an understanding of both temporal and spectral patterns that define musical styles. Traditional genre classification methods have largely depended on handcrafted audio features, which capture some, but not all, aspects of musical structure  \cite{zhang2014deep} \cite{mayer2008combination} \cite{fu2010survey}. Recently, deep learning models have demonstrated promising improvements by leveraging large datasets and data-driven feature extraction \cite{choi2017convolutional} \cite{dong2018convolutional}. However, these models often fail to account for the temporal significance of musical segments, an element critical for accurately capturing genre characteristics \cite{dong2018convolutional} \cite{costa2017evaluation}. Furthermore, existing models that incorporate temporal dynamics tend to be computationally complex \cite{zhuang2020music} and fail to extract genre-specific signatures from music pieces, which could be leveraged for data compression and more efficient recommendation systems. \cite{dai2016long} \cite{yu2020deep} \cite{chen2024hybrid}

Our motivation stems from the need to uncover temporal patterns within music that are pivotal for defining genres, which, in turn, enable us to understand "signature moments" in a track. These moments often resonate strongly with listeners, making them fundamental to both genre recognition and the listener experience. In this study, we propose an attention-guided model that processes sequences of spectrogram images, prioritizing the temporally significant segments to generate a distinctive signature for each music genre. By analyzing these temporal signatures, we gain insights into genre proximity and contrast, enriching the music recommendation landscape \cite{schedl2019deep}.

Our contributions are threefold: (1) We develop a CNN-based attention model that identifies the most impactful temporal segments for genre classification, \cite{costa2017evaluation} (2) we utilize these insights to establish genre signatures that can improve recommendation algorithms and deepen genre relationships, and (3) we map these findings to align with intuitive human perceptions of genre, highlighting potential applications in music recommendation and generation.

\section{Proposed Method}

Our approach integrates Convolutional Neural Networks (CNNs) with an attention mechanism to classify music genres by capturing the temporal significance of spectrogram segments. By leveraging CNNs for spatial feature extraction and multi-head attention for temporal focus, we obtain a model capable of identifying genre-defining “signature moments” in a music piece.

\subsection{Model Architecture} The architecture consists of three main components: a CNN to extract spatial features from spectrogram images, an attention mechanism to capture temporal dependencies, and a classification layer for genre prediction. Each spectrogram sequence is processed to identify impactful segments, enhancing classification accuracy and providing interpretability. The CNN processes each spectrogram sequence through a series of convolutional layers, extracting spatial features from each frame \cite{dong2018convolutional}. A TimeDistributed layer then maintains the sequence structure as these features are passed through it. A multi-head attention mechanism is subsequently applied to the sequence of feature tokens, enabling the model to focus on key temporal segments that are crucial for genre classification \cite{yu2020deep} \cite{xie2024music}. Finally, the attention-weighted sequence representation is passed through a fully connected layer for the final genre prediction \cite{costa2017evaluation}.
\end{multicols}
\begin{center} \includegraphics[width=0.8\textwidth]{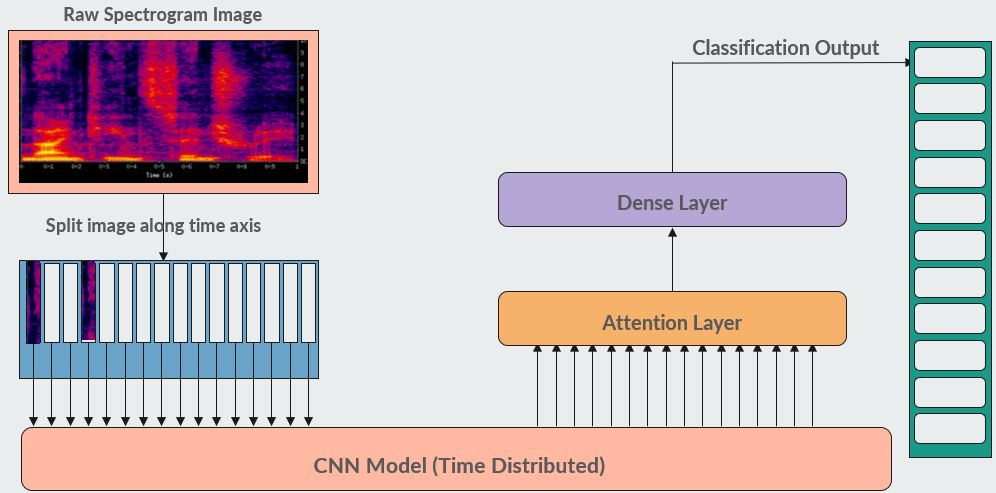}\ \text{Figure 1: Model Architecture} 
\end{center}

\begin{multicols}{2} To the sequence of feature tokens, enabling the model to focus on key temporal segments that are crucial for genre classification. Finally, the attention-weighted sequence representation is passed through a fully connected layer for the final genre prediction.

\subsection{Datasets} We used the GTZAN dataset for training and evaluation, which includes 1,000 30-second audio tracks spanning 10 genres. Each track is represented by grayscale spectrograms, which are then cut along the x-axis (or the time axis) to form what we will call "tokens." The size of each of these cropped tokens is 217x45, which corresponds to exactly 4 seconds of the piece. There exists an overlap of around 1 second. That is, the first token will be from the start of the piece to 4 seconds. The second token will be from 3 seconds to 7 seconds, and so on to form a total of 10 tokens of the piece.

\subsection{Methodology and Advantages} The model architecture is designed to leverage CNNs for spatial feature extraction from each spectrogram, followed by a multi-head attention mechanism to capture the most important temporal segments. This structure offers several advantages. First, by capturing temporal dependencies, the model identifies genre-specific patterns over time, which are essential for genre differentiation \cite{zhuang2020music}. The segmented spectrograms allow for enhanced contextual understanding, contributing to a nuanced temporal representation that improves classification accuracy \cite{choi2017convolutional}. The CNN layers facilitate robustness to noise, enabling the model to extract meaningful features even in less ideal conditions \cite{clark2012music}.

A key benefit of this model is its ability to extract genre-specific "signature moments" by highlighting critical spectrogram segments, which can reduce computational requirements for recommendation systems and help listeners and musicians alike understand the unique characteristics of genres, such as reggae’s rhythmic style or jazz’s improvisational nature \cite{corrigall2015liking}. Additionally, the model’s use of attention scores enables interpretability, allowing insight into which spectrogram segments most influence the classification, thus providing a transparent decision-making process \cite{nanni2016combining}. \cite{pachet2000taxonomy}

\section{Experiments and Results}

The experiments conducted aimed to evaluate the classification accuracy of the proposed CNN with attention mechanism on the task of music genre classification. Several analyses were performed, including confusion matrix analysis, PCA-based visualizations, and attention weight analysis to interpret which aspects of the music contributed most significantly to genre classification. Below, we detail the experimental setup, results, and findings.

\subsection{Classification Accuracy and Confusion Matrix}
The model was trained using 6-fold cross-validation, and the classification accuracy was evaluated using a confusion matrix. This matrix highlights which genres the model is able to distinguish well and where confusion arises. Notably:
\begin{itemize}
    \item \textbf{Misclassification Patterns:} Genres such as \textit{blues} and \textit{country} showed some confusion due to their shared musical roots. Similarly, \textit{hiphop} and \textit{reggae} were often misclassified due to rhythmic and production similarities. 
    \item \textbf{High Accuracy Genres:} Certain genres, such as \textit{classical} and \textit{jazz}, showed high classification accuracy, likely because of their distinct lack of percussion and rhythmic structure, which sets them apart from modern genres like \textit{disco} or \textit{pop}.
\begin{center}
    \includegraphics[width=0.4\textwidth]{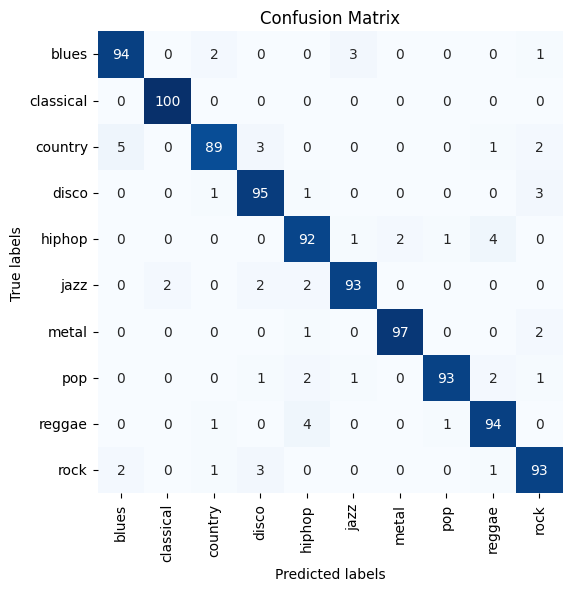}\\ 
    \text{Figure 2: Confusion Matrix}
\end{center}
\end{itemize}

This analysis suggests that genres with strong rhythmic or production similarities are more challenging for the model to distinguish, whereas genres with unique structural features, such as \textit{classical} and \textit{jazz}, are easier to classify.

\subsection{Genre Embedding based Experiments}
In this experiment, we aimed to analyze the relationships between musical genres based on the individual characteristics of each music piece. Each audio piece was encoded using a convolutional neural network (CNN), and the resulting embeddings were weighted according to the softmax of the corresponding attention scores. This process allowed us to assign greater significance to parts of the audio piece that were deemed more important by the attention mechanism.

Subsequently, all embeddings within the same genre were averaged to generate a comprehensive genre encoding. This approach facilitated the capture of the inherent features shared among pieces of the same genre. 

\subsubsection{PCA Analysis of Genre Distribution}
Principal Component Analysis (PCA) was performed on the genre embeddings extracted from the CNN to reduce the dimensionality and visualize the distribution of genres in 2D and 3D space. The PCA plots provide insights into how genres cluster based on learned features:
\begin{itemize}
    \item \textbf{2D Visualization of Genre Clustering:} Certain genres, such as \textit{classical} and \textit{jazz}, are far from the rest of the clusters, indicating their uniqueness in terms of instrumentation and structure. Conversely, genres like \textit{disco} and \textit{pop} are more closely clustered, reflecting their shared upbeat tempo and electronic instrumentation.
    \item \textbf{3D Exploration of Genre Relationships} In the 3D plot, certain genres concentrate along specific axes, suggesting that some PCA components may represent rhythmic elements or electronic instrumentation. For example, \textit{rock}, \textit{disco}, and \textit{hiphop} might all share rhythmic components, placing them on similar axes, despite other differences in tempo and melody.
\end{itemize}

The PCA analysis highlights the relationships between genres, allowing us to visualize how closely related they are in terms of the features learned by the model.

\subsubsection{PCA-Based Genre Relationships}
The PCA-Based Genre Relationship Equations suggest a mathematical relationship between different genres. These equations arise by computing combinations of genre embeddings that satisfy the equation (A - B + C = D, where A, B, C, and D are the embeddings of four different genres), upto a threshold error. \\These equations describe how the features of one genre can be modified to resemble another. Some notable relationships include:
\begin{itemize}
    \item \textbf{blues - country + disco = rock:} This suggests that by subtracting "country" features from \textit{blues} and adding \textit{disco}'s upbeat tempo, we get \textit{rock}. This reflects the evolution of rock music from its blues origins, combined with the rhythm of more modern genres.
    \item \textbf{disco - metal + rock = pop:} Here, \textit{disco}'s rhythm and energy, when combined with \textit{rock}'s structure and subtracting \textit{metal}'s aggression, result in \textit{pop} music. This aligns with how pop music often incorporates elements from both disco and rock.
    \item \textbf{disco - reggae + pop = rock:}  Captures rock as a blend of disco’s rhythm, minus reggae’s slower pace, and infused with pop's accessible elements, creating an \textit{energetic genre.}
\end{itemize}

\begin{center}
    \includegraphics[width=0.3\textwidth]{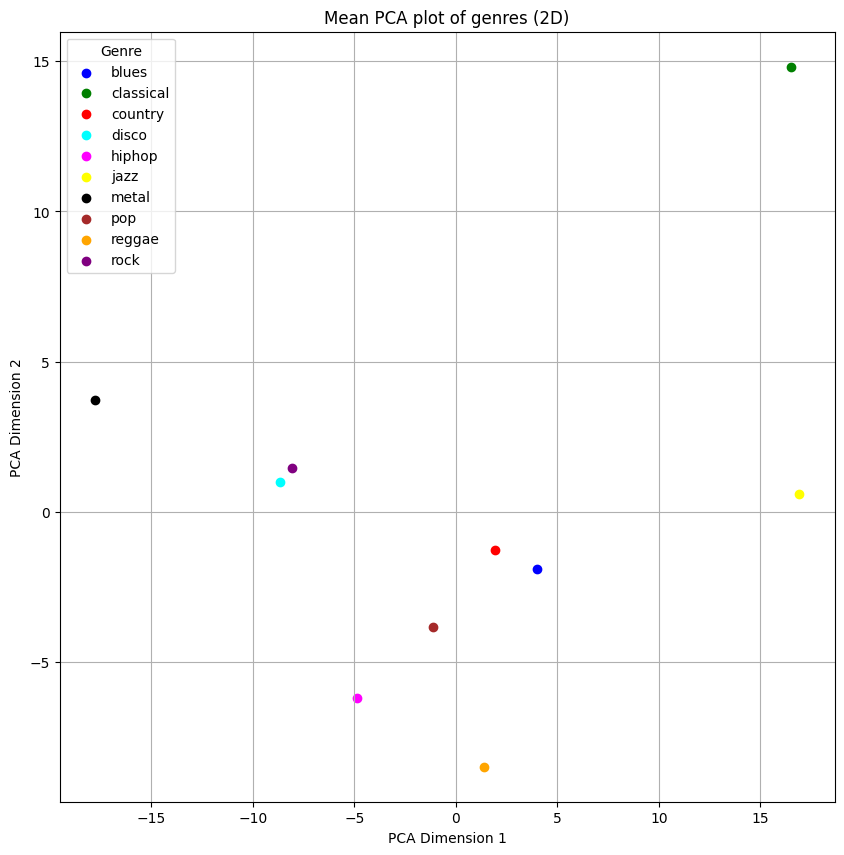}\\ 
    \text{Figure 3: 2D PCA Plot}
\end{center}
\begin{center}
    \includegraphics[width=0.3\textwidth]{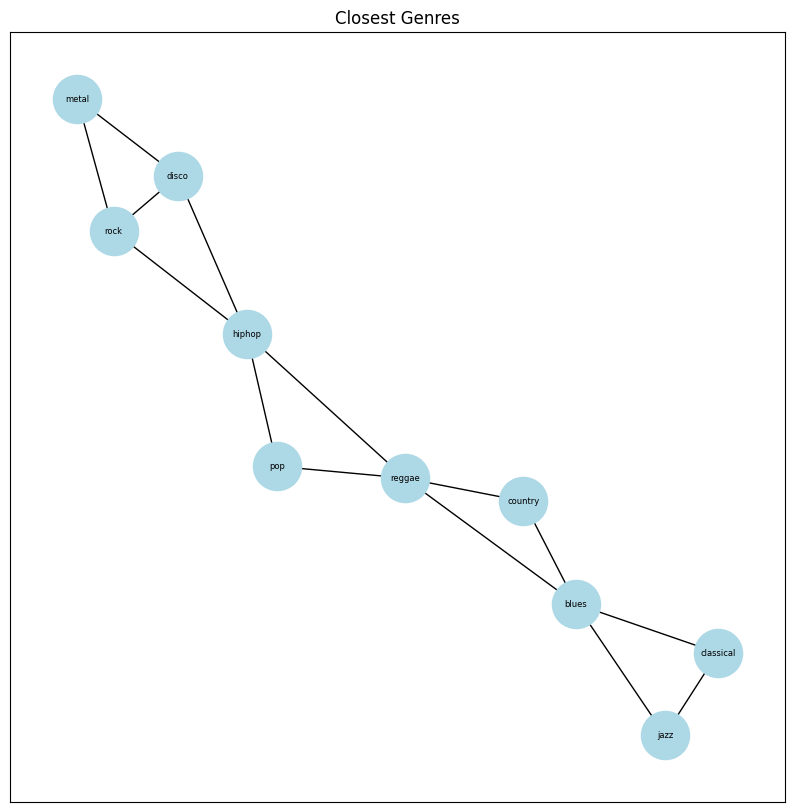}\\ 
    \text{Figure 4: Closest Genres}
\end{center}
These equations provide a fascinating mathematical insight into how genres can be transformed into one another by adjusting specific musical features, highlighting the shared characteristics between different genres.

\subsection{Comparative Analysis of Musical Genres}
\subsubsection{Experiment Definition}
To determine the closest genres to each genre, we applied a 2-nearest neighbors (2-NN) algorithm on these genre embeddings. This method identified the two most similar genres for each genre based on the calculated distances in the encoding space, thereby enabling a quantitative assessment of genre similarity.

\subsubsection{Observations and Analysis}
\begin{itemize}
    \item \textbf{Blues:} Closest to \textit{country} and \textit{reggae}, likely due to the shared slow, soulful elements found in both genres.
    \item \textbf{Classical:} Closest to \textit{jazz} and \textit{blues}, which makes sense due to the instrumental complexity and
    \begin{center}
    \includegraphics[width=0.3\textwidth]{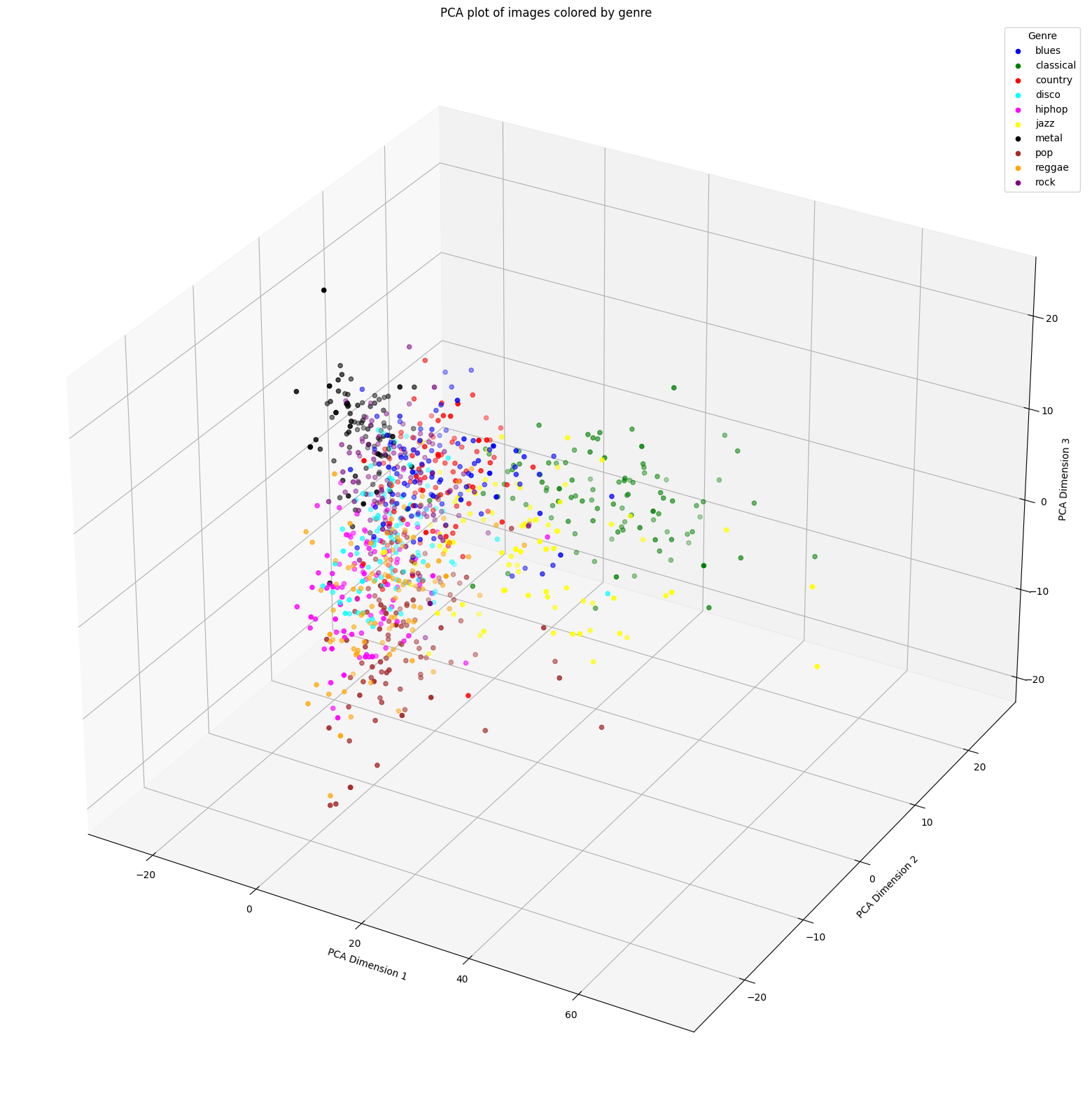}\\ 
    \text{Figure 5: 3D PCA Plot}
    \end{center}
    \begin{center}
    \includegraphics[width=0.3\textwidth]{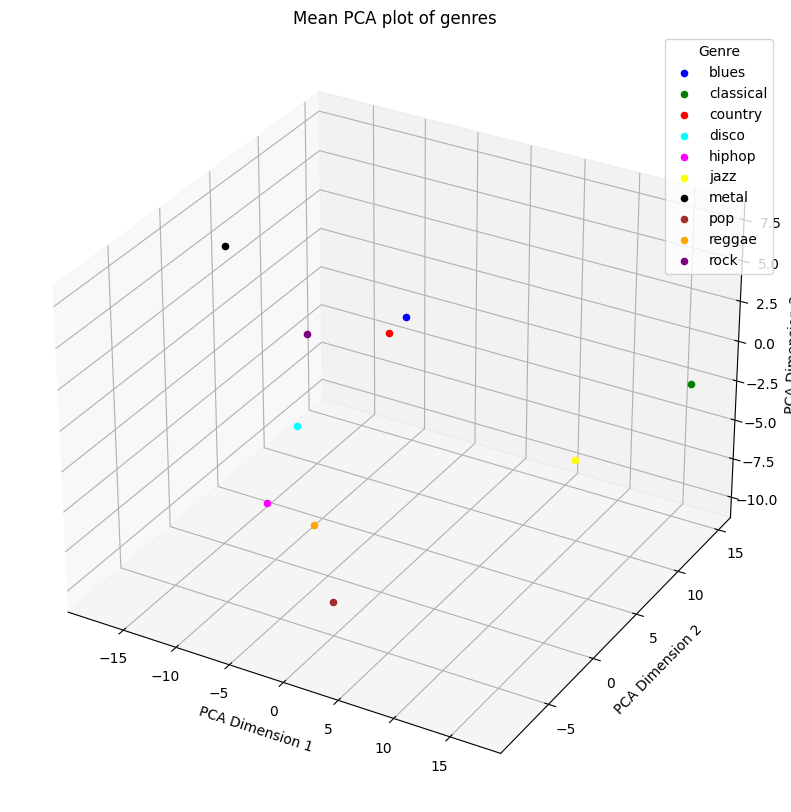}\\ 
    \text{Figure 6: Additional 3D PCA Plot}
    \end{center}
    improvisational nature of jazz and the emotional depth of blues.
    \item \textbf{Metal:} Farthest from \textit{classical} and \textit{jazz}, as these genres are fundamentally different in terms of instrumentation, structure, and rhythm.
\end{itemize}

\subsection{Attention Score Analysis Across Music Genres}
\subsubsection{Experiment Definition}In this experiment, we analyzed attention scores for five random samples from each genre, revealing several significant observations regarding musical elements. These "most important" time steps of the music piece were then analyzed by listening to those time steps, figuring out patterns, and constructing intuitive hypotheses.

\subsubsection{Observations and Analysis}The model highlighted that the first occurrence of similar patterns within a piece is the most important, suggesting effective compression of the music with minimal information loss. In blues, rhythmic sections and classic bending guitar styles were identified as critical elements, reflecting their unique signature within the genre. 

The conclusion of pieces, particularly in classical and blues, was also significant. The avoidance of abrupt endings in the dataset emphasizes the importance of a structured conclusion in genre identity.

For country music, attention focused on instruments like the harmonica and fiddle, along with harmonizations that are characteristic of the genre. In jazz, "out-of-scale" notes received high attention, highlighting the genre's improvisational nature.

In metal, significant attention was given to intense sections with drums, guitars, and growls, as well as isolated guitar solos with heavy drumming, indicating dynamic contrasts. For pop, the chorus and pre-chorus were marked as important, consistent with the genre's emphasis on catchy hooks.

Reggae's unique rhythmic patterns were evident in the significance of instrumental sections, while in rock, distinctive drum fills and the presence of the flute from the late 60s emerged as notable features.

These findings illustrate the model's ability to discern genre-specific characteristics and provide insights into the relationships between musical elements and genre classification.

\subsection{Signature Analysis for Music Recommendation}
\subsubsection{Experiment Definition}This experiment aims to identify the "signature" of a random music piece and compare it to the embeddings of signatures from other pieces, thereby determining the top five most similar recommendations according to the model. The objective is to evaluate the effectiveness of the signature in music recommendation.
The most similar signatures are extracted by computing the trained CNN Model encoding and performing K-Nearest Neighbors on the same. 

\subsubsection{Observations and Analysis}In the classical genre, a purely string instrumental piece was recommended alongside another string-only composition. Additionally, pieces with similar scales and frequencies from classical and blues were identified as closely related. The model also associated slow, melancholy classical pieces with similar jazz compositions, indicating shared emotional characteristics.

For disco music, pieces featuring significant rock elements, such as guitar solos and synthesizers, were recommended alongside rock and metal tracks that shared similar time signatures. The recommended metal pieces exhibited more rock characteristics than typical metal elements, all adhering to a 3/4 time signature, akin to the disco tracks.

In the metal genre, all suggested pieces maintained a strong metal identity, reflecting the genre's distinctive sound profile. 

For blues, the model predominantly suggested other blues or jazz pieces, which shared similar scales, time signatures, and smooth characteristics, reinforcing the interconnectedness of these genres. 

In pop music, slow pop tracks were recommended alongside other slow pieces, while fast pop songs were matched with similarly fast tracks, highlighting the consistency in vocal texture across the suggested pieces.

These observations illustrate the model's capability to identify and recommend pieces based on genre-specific signatures, thereby enhancing music recommendation accuracy.

\section{Conclusion}
This study explored the application of convolutional neural networks and attention mechanisms in the domain of music genre classification and recommendation. Through rigorous experimentation, we demonstrated that our proposed model effectively captures the intricate features of musical pieces, leading to meaningful genre representations and insights into musical similarities.

Our findings reveal that the attention mechanism not only highlights crucial segments within each piece but also enhances the model's interpretability by providing visibility into the significance of specific audio features. The genre-based analysis confirmed human intuitions regarding musical relationships, with genres exhibiting shared elements clustering together.

Furthermore, the ability to derive a "signature" from individual pieces and recommend similar works underscores the model's potential in music recommendation systems. Overall, this research contributes to the understanding of how deep learning techniques can be utilized to advance music classification and recommendation, paving the way for future studies to build upon these findings and further refine genre recognition models.
\end{multicols}
\newpage
\begin{center}
    \includegraphics[width=1\textwidth]{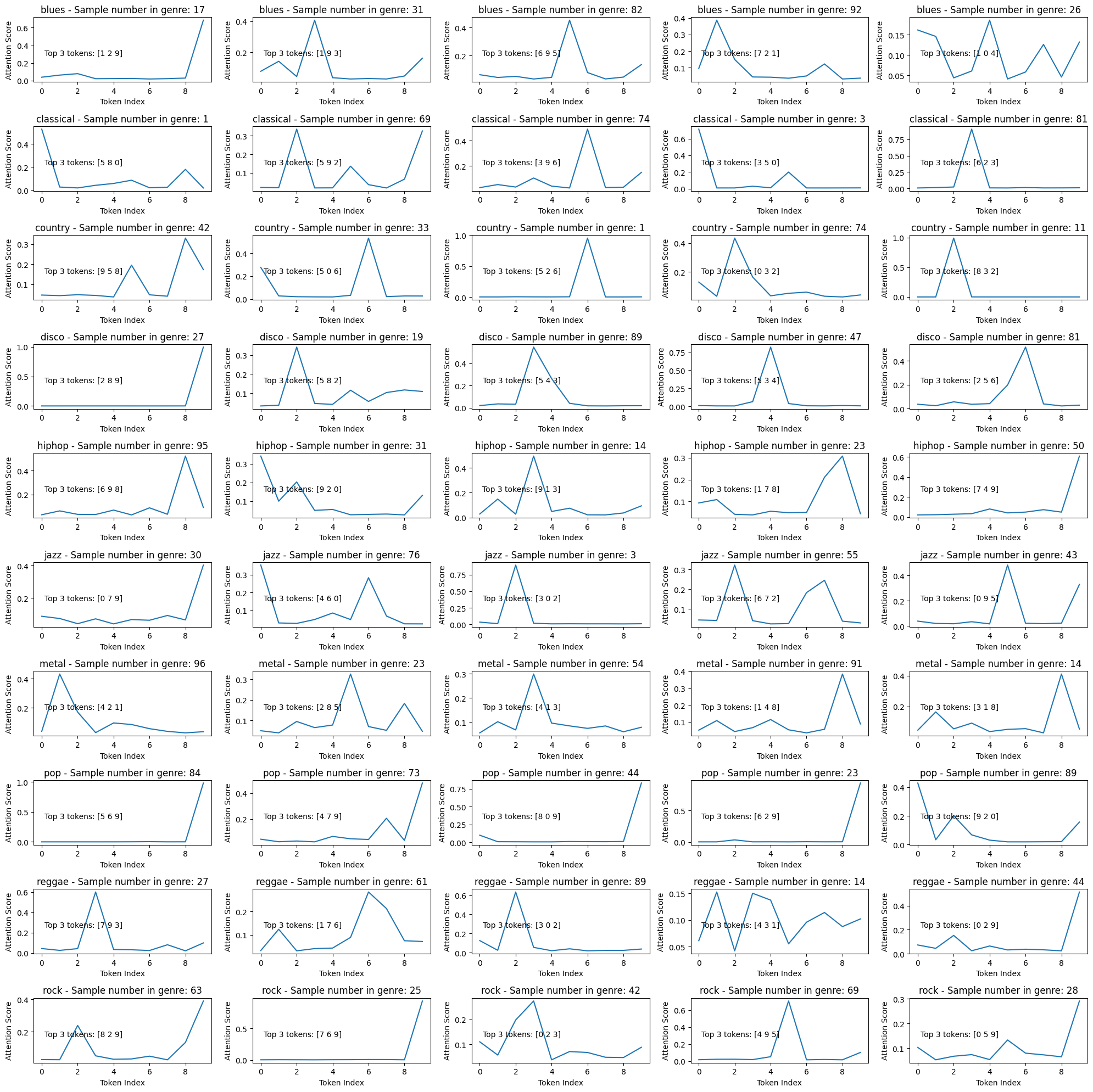}\\ 
    \text{Figure 7: Attention Weights Analysis}
    
\end{center}
\newpage
\begin{multicols}{2}

\end{multicols}


\begin{thebibliography}{10}

\bibitem{chen2024hybrid}
Jiyang Chen, Xiaohong Ma, Shikuan Li, Sile Ma, Zhizheng Zhang, and Xiaojing Ma.
\newblock A hybrid parallel computing architecture based on cnn and transformer for music genre classification.
\newblock {\em Electronics}, 13(16):3313, 2024.

\bibitem{choi2017convolutional}
Keunwoo Choi, Gy{\"o}rgy Fazekas, Mark Sandler, and Kyunghyun Cho.
\newblock Convolutional recurrent neural networks for music classification.
\newblock In {\em 2017 IEEE International conference on acoustics, speech and signal processing (ICASSP)}, pages 2392--2396. IEEE, 2017.

\bibitem{clark2012music}
Sam Clark, Danny Park, and Adrien Guerard.
\newblock Music genre classification using machine learning techniques, 2012.

\bibitem{corrigall2015liking}
Kathleen~A Corrigall and E~Glenn Schellenberg.
\newblock Liking music: Genres, contextual factors, and individual differences.
\newblock {\em Art, aesthetics, and the brain}, 15(1), 2015.

\bibitem{costa2017evaluation}
Yandre~MG Costa, Luiz~S Oliveira, and Carlos~N Silla~Jr.
\newblock An evaluation of convolutional neural networks for music classification using spectrograms.
\newblock {\em Applied soft computing}, 52:28--38, 2017.

\bibitem{dai2016long}
Jia Dai, Shan Liang, Wei Xue, Chongjia Ni, and Wenju Liu.
\newblock Long short-term memory recurrent neural network based segment features for music genre classification.
\newblock In {\em 2016 10th International Symposium on Chinese Spoken Language Processing (ISCSLP)}, pages 1--5. IEEE, 2016.

\bibitem{dong2018convolutional}
Mingwen Dong.
\newblock Convolutional neural network achieves human-level accuracy in music genre classification.
\newblock {\em arXiv preprint arXiv:1802.09697}, 2018.

\bibitem{fu2010survey}
Zhouyu Fu, Guojun Lu, Kai~Ming Ting, and Dengsheng Zhang.
\newblock A survey of audio-based music classification and annotation.
\newblock {\em IEEE transactions on multimedia}, 13(2):303--319, 2010.

\bibitem{mayer2008combination}
Rudolf Mayer, Robert Neumayer, and Andreas Rauber.
\newblock Combination of audio and lyrics features for genre classification in digital audio collections.
\newblock In {\em Proceedings of the 16th ACM international conference on Multimedia}, pages 159--168, 2008.

\bibitem{moore2012introduction}
Brian~CJ Moore.
\newblock {\em An introduction to the psychology of hearing}.
\newblock Brill, 2012.

\bibitem{nanni2016combining}
Loris Nanni, Yandre~MG Costa, Alessandra Lumini, Moo~Young Kim, and Seung~Ryul Baek.
\newblock Combining visual and acoustic features for music genre classification.
\newblock {\em Expert Systems with Applications}, 45:108--117, 2016.

\bibitem{pachet2000taxonomy}
Fran{\c{c}}ois Pachet, Daniel Cazaly, et~al.
\newblock A taxonomy of musical genres.
\newblock In {\em RIAO}, volume~2, pages 1238--1245. Citeseer, 2000.

\bibitem{schedl2019deep}
Markus Schedl.
\newblock Deep learning in music recommendation systems.
\newblock {\em Frontiers in Applied Mathematics and Statistics}, 5:457883, 2019.

\bibitem{xie2024music}
Changjiang Xie, Huazhu Song, Hao Zhu, Kaituo Mi, Zhouhan Li, Yi~Zhang, Jiawen Cheng, Honglin Zhou, Renjie Li, and Haofeng Cai.
\newblock Music genre classification based on res-gated cnn and attention mechanism.
\newblock {\em Multimedia Tools and Applications}, 83(5):13527--13542, 2024.

\bibitem{yu2020deep}
Yang Yu, Sen Luo, Shenglan Liu, Hong Qiao, Yang Liu, and Lin Feng.
\newblock Deep attention based music genre classification.
\newblock {\em Neurocomputing}, 372:84--91, 2020.

\bibitem{zhang2014deep}
Chiyuan Zhang, Georgios Evangelopoulos, Stephen Voinea, Lorenzo Rosasco, and Tomaso Poggio.
\newblock A deep representation for invariance and music classification.
\newblock In {\em 2014 IEEE international conference on acoustics, speech and signal processing (ICASSP)}, pages 6984--6988. IEEE, 2014.

\bibitem{zhuang2020music}
Yingying Zhuang, Yuezhang Chen, and Jie Zheng.
\newblock Music genre classification with transformer classifier.
\newblock In {\em Proceedings of the 2020 4th international conference on digital signal processing}, pages 155--159, 2020.

\end{thebibliography}
\end{document}